\documentclass[12pt]{article}
\usepackage[utf8]{inputenc}
\usepackage{amsfonts}
\usepackage{amsmath}
\usepackage{amssymb}
\usepackage{graphicx}
\usepackage{color}
\usepackage[left=1cm,top=1cm,bottom=1cm,right=1cm,nohead,nofoot]{geometry}

\def \be {\begin{equation}}
\def \ee {\end{equation}}
\def \bea {\begin{eqnarray}}
\def \eea {\end{eqnarray}}
\def \nn {\nonumber}

\def \rr {\raise.35ex\hbox{\small $\prime$}\kern-.17em{\mbox{\large $\imath$}}}
\def \del {\partial}
\def \dels {\partial\kern-.6em /\kern.1em}
\def \As {{A\kern-.5em / \kern.5em}}
\def \Ds {D\kern-.7em / \kern.5em}
\def \a {\alpha}

\def \eps {\epsilon}

\def \ks {k\kern-.5em /}
\def \ls {l\kern-.5em /}

\def \om {\omega}

\def \sgn {\mbox{\small sgn}}

\def \da {\dot{\alpha}}
\def \db {\dot{\beta}}
\def \dg {\dot{\gamma}}

\reversemarginpar
\newcommand{\hide}[1]{}

\begin{document}
\begin{titlepage}

\begin{center}

\hfill
\vskip .2in

\textbf{\LARGE
Dimensional Reduction of the Generalized DBI
\vskip.5cm
}

\vskip .5in
{\large
Jun-Kai Ho$\,{}^{a,c}$\footnote{e-mail address: junkai125@gmail.com} and
Chen-Te Ma$^{a,b}$\footnote{e-mail address: yefgst@gmail.com}\\
\vskip 3mm
}
{\sl
${}^a$
Department of Physics, Center for Theoretical Sciences,\\
${}^b$
Center for Advanced Study in Theoretical Sciences, 
National Taiwan University, Taipei 10617, Taiwan,
R.O.C.\\
${}^c$
Department of Physics, Brown University, Box 1843 Providence, RI 02912-1843, USA.}\\
\vskip 3mm
\vspace{60pt}
\end{center}
\begin{abstract}
We study the generalized Dirac-Born-Infeld (DBI) action, which describes a $q$-brane ending on a $p$-brane with a ($q$+1)-form background. This action has the equivalent descriptions in commutative and non-commutative settings, which can be shown from the generalized metric and Nambu-Sigma model. We mainly discuss the dimensional reduction of the generalized DBI at the massless level on the flat spacetime and constant antisymmetric background in the case of flat spacetime, constant antisymmetric background and the gauge potential vanishes for all time-like components. In the case of $q=2$, we can do the dimensional reduction to get the DBI theory. We also try to extend this theory by including a one-form gauge potential.

\end{abstract}

\end{titlepage}

\section{Introduction}
\label{1}
In string theory, the T-duality shows the equivalence of two theories that look different under the exchange of a radii R and $\alpha^{\prime}$/R. For the closed strings \cite{Zwiebach:1992ie, Saadi:1989tb}, the T-duality of closed strings exchanges winding and momentum modes. In the case of open strings, the T-duality of open strings exchanges the Dirichlet and Neumann boundary conditions. The studies of the T-duality pave the way for the unified theory. One way is to study the low energy effective action. It is well-known that the DBI model can be derived from the one-loop $\beta$ function. The low-energy effective field theory (higher derivative gravity \cite{Zwiebach:1985uq} or non-local theory) can be directly found from the one-loop $\beta$ function. Nevertheless, we encounter the non-geometry or T-fold problem in the massless closed string theory. This problem is generic and unavoidable in string theory. To solve this problem, we need to construct geometric languages to endow string theory with a global geometry. Double field theory \cite{Hull:2009mi, Hull:2009zb, Hohm:2010jy, Hohm:2010pp, Lee:2013hma, Tseytlin:1990va, Tseytlin:1990nb, Copland:2011wx, Berman:2007yf, Berman:2007xn, Avramis:2009xi, Berkeley:2014nza, Duff:1989tf, Siegel:1993xq, Siegel:1993th, Siegel:1993bj} and generalized geometry \cite{Gualtieri:2003dx, Hitchin:2004ut, Cavalcanti:2011wu} are two typical examples. This new \lq\lq{}stringy geometry\rq\rq{} \cite{Hohm:2011dv, Hohm:2010xe} geometrize the non-geometric flux and find the 10-dimensional supergravity with the non-geometric flux as shown in \cite{Aldazabal:2011nj, Andriot:2012an}. At the current stage, the most non-trivial problem is that the double field theory needs the section conditions to have gauge invariance. The approaches of relaxing the section conditions can be found in \cite{Geissbuhler:2013uka, Grana:2012rr, Ma:2014ala}. The extension of the $\alpha^{\prime}$ correction explores a geometric way to find the T-duality with $\alpha^{\prime}$ correction \cite{Hohm:2013jaa, Bedoya:2014pma}. The geometric structure can also be extended to the 11-dimensional supergravity \cite{Berman:2010is, Hohm:2013vpa, Hohm:2013uia, Hohm:2014fxa, Berman:2012vc, Hohm:2013pua}. Some recent good reviews can be found in \cite{Hohm:2013bwa, Aldazabal:2013sca, Berman:2013eva}.

To get a geometric picture of brane theory, a combination of non-commutative gauge theory and the generalized geometry is necessary. The non-commutative gauge theory of the D-brane is already well-known, but the non-commutative gauge theory of the M-brane is still not completely understood \cite{Ho:2008nn, Ho:2012dn}. Recently, theories based on the equivalence between the commutative and non-commutative gauge theories are found. The theories are the Nambu-Sigma model and generalized DBI model \cite{Jurco:2012yv, Schupp:2012nq}. The non-commutative geometry is encoded in the generalized metric, which is an ingredient of the generalized geometry. Although they do not use the full language of the generalized geometry \cite{Jurco:2013upa, Jurco:2014sda}, they found the evidence for the DBI-like M2-M5 system \cite{Cederwall:1997gg}.

The main task of this paper is to calculate the dimensional reduction of the generalized DBI theory at the massless level. We perform the dimensional reduction from a $(q+1)$-brane ending on a $(p+1)$-brane to a $q$-brane ending on a $p$-brane. We consider flat spacetime, constant antisymmetric background field and the $(q+1)$-form gauge field only exists in $(q+1)$-dimensional worldvolume directions (no time direction) in $q$-$p$ system. The non-trivial result of this theory is that the appearance of the 2$(q+1)$-th root, which can be shown by the equivalence between the commutative and non-commutative descriptions, is robust against the dimensional reduction. The most interesting study is the system of a 2-brane ending on a  5-brane. The system can be reduced to a 1-brane ending on a 4-brane by the dimensional reduction. This shows that the system of a 2-brane ending on a 5-brane can be reduced to the DBI theory in our simple consideration. Finally, we discuss the possibility of adding one-form gauge field in the generalized DBI theory. We can include one-form gauge field up to $H^2$ in principle, and the calculation also demonstrates the potential to extend the generalized DBI theory with different field contents. This study should give the simplest understood of the higher-form fields although it is not a general consideration. Our study on the generalized DBI theory should motivate interest of duality structure in higher dimensions.

The plan of this paper is to first review the generalized DBI theory in Sec. \ref{2}.
Then we discuss the dimensional reduction without scalar fields in Sec. \ref{3} and dimensional reduction with scalar fields in Sec. \ref{4}.
Finally, we conclude in Sec. \ref{5}. We also provide the detailed calculation in \ref{app1} and \ref{app2}.

\section{Review of the Generalized DBI}
\label{2}
In this section, we follow \cite{Jurco:2012yv, Schupp:2012nq} to review the generalized DBI theory. First of all, we show the closed-open string relations from the string sigma model. Secondly, we generalize the Poisson-Sigma model to Nambu-Sigma model. We obtain the generalized closed-open relations from the Nambu-Sigma model. Then we introduce the membrane action, which is equivalent to the Nambu-Sigma model under the gauge fixing. In the end of this section, we use the generalized closed-open relations to construct an action.

We define our notations as follow. We denote the worldvolume directions from $A$ to $H$ and indicate the transverse directions from $I$ to $Z$. The index $a$=$1, 2, \cdots, p$ are reserved for the spatial components of worldvolume coordinates (We denote them from $a$ to $h$.), while the Greek letters $\mu, \nu, \rho, \sigma$=0, 1, $\cdots$, $D-1$ denote the target space indices and $w$=0, 1 denote the world-sheet index. In addition, we use $i$, $j$ to label the antisymmetric indices, $i=(i_1, i_2, \cdots, i_{r})$ with $0\leq i_1<i_2<\cdots<i_r\leq (r+1)$, where $r$ is the dimension of $i$. 

\subsection{Closed-Open Relations}
We first introduce the action of the Poisson-Sigma model 
\begin{equation}
S_P = \int_\Sigma\ \left( A_\mu \wedge dX^\mu - \frac{1}{2} \Pi^{\mu\nu} A_\mu \wedge A_\nu\right), \qquad \Pi \equiv \frac{1}{2} \Pi^{\mu\nu}(X)\partial_\mu \wedge \partial_\nu ,
\end{equation}
where $X:\Sigma\rightarrow M$, $\Sigma$ is the two dimensional world-sheet and $M$ is the target space manifold. The one-form field $A(\sigma)$ is on $\Sigma$ and $\Pi$ is an antisymmetric tensor. From the equations of motion
\bea
dX^{\mu}-\Pi^{\mu\nu}A_{\nu}=0,\qquad dA_{\mu}+\frac{1}{2}\partial_{\mu}\Pi^{\nu\rho}A_{\nu}\wedge A_{\rho}=0,
\eea
we show that the bi-vector $\Pi$ satisfies the Jacobi identity. These are the equations of motion for $A_\mu$ and $X^\mu$, respectively. We can add a metric term in the Poisson-Sigma model to obtain the non-topological generalized Poisson-Sigma model
\begin{equation} \label{genPoissonsigma}
S_P = \int_\Sigma\ \left( A_\mu \wedge dX^\mu - \frac{1}{2} \Pi^{\mu\nu} A_\mu \wedge A_\nu - \frac{1}{2} (G^{-1})^{\mu\nu} A_\mu \wedge *A_\nu \right) ,
\end{equation}
where $*A_{\nu}$ is the Hodge dual of $A_{\nu}$. The signature of the world-sheet is $(-,+)$ and volume form $d^2 \sigma \equiv d\sigma^0 \wedge d\sigma^1$. The $A_\mu \equiv A_{\mu w}(\sigma) d\sigma^w$ is an auxiliary field. By using the equation of motion of $A_\mu$, the action \eqref{genPoissonsigma} can be rewritten as the string sigma model action,
\begin{equation}
S_S = -\int_\Sigma\ \frac{1}{2} \left(g_{\mu\nu} dX^{\mu} \wedge *dX^{\nu} + B_{\mu\nu} dX^{\mu} \wedge dX^{\nu}\right) ,
\end{equation}
where the $g$ and $B$ are defined by the closed-open string relations 
\begin{equation} \label{close-open-string}
\frac{1}{g+B} = G^{-1} + \Pi  , \Rightarrow\qquad G=g-Bg^{-1}B,\qquad\Pi=-G^{-1}Bg^{-1}=-g^{-1}BG^{-1}.
\end{equation}  
The action \eqref{genPoissonsigma} can also be rewritten in terms of the components of $\eta_\mu \equiv  -A_{\mu 1}(\sigma)$ and $\tilde \eta_\nu \equiv A_{\nu 0}(\sigma)$, the action is
\bea \label{poissonsigma_eta}
    S_P = \int d^2\sigma\ \bigg[ -\frac{1}{2}
    (G^{-1})^{\mu\nu} \eta_\mu \eta_\nu + \frac{1}{2} (G^{-1})^{\mu\nu} \tilde{\eta}_\mu \tilde{\eta}_\nu 
    + \eta_\mu \partial_0 X^\mu + \tilde{\eta}_\mu \partial_1 X^\mu - \Pi^{\mu \nu} \eta_\mu \tilde{\eta}_\nu\bigg] .
\nn\\
\eea
 We can use matrix notation to rewrite the action by using
 \bea
 \eta\equiv\eta_{\mu},\qquad\tilde{\eta}\equiv\tilde{\eta}_{\nu}, \qquad G\equiv G_{\mu\nu}, \qquad X\equiv X^{\mu}, \qquad\Pi\equiv\Pi^{\mu\nu}.
 \eea
 The action becomes
 \bea
     S_P = \int d^2\sigma\ \bigg( -\frac{1}{2}
    \eta^T G^{-1}\eta + \frac{1}{2} \tilde{\eta}^TG^{-1}\tilde{\eta} 
    + \partial_0 X^T\eta+ \partial_1 X^T\tilde{\eta}  - \eta^T\Pi\tilde{\eta}\bigg) ,
 \eea
where the superscript $T$ indicates transpose of matrix. By using the matrix notation, it is easier to generalize the Poisson-Sigma model.

\subsection{Generalized Closed-Open Relations}
The Nambu-Sigma model is a generalization of the Poisson-Sigma model. The action is given by
\begin{equation} \label{NambuSigma}
    S_N = \int d^{q+1}\sigma\ \bigg( -\frac{1}{2}
    \eta^T G^{-1}\eta+ \frac{1}{2} \tilde{\eta}^T\tilde{G}^{-1} \tilde{\eta} + \partial_0 X^T\eta +  \widetilde{\partial\! X}^T\tilde{\eta} - \eta\Pi^T \tilde{\eta}\bigg) ,
\end{equation}
where
\begin{equation}\label{gtilde}
    \tilde G_{ij} = \sum_{\pi } \sgn(\pi) G_{i_{\pi(1)} j_1} \cdots  G_{i_{\pi(p)} j_q}
\end{equation}
with a permutation $\pi$. The antisymmetric product of partial derivatives is defined as 
\begin{equation}
\widetilde{\partial\! X}^i \equiv \sum_{a_1, \ldots, a_q =1}^q \epsilon^{a_1 a_2 \ldots a_q} \partial_{a_1} X^{i_1} \cdots \partial_{a_q} X^{i_q},
\end{equation}
where $0\leq i_1 < \cdots < i_q\leq (q+1)$.
There are two types of the metrics $G$ and $\tilde{G}$, auxiliary fields $\eta$ and $\tilde{\eta}$, and an antisymmetric $(q+1)$-form tensor $\Pi$.
We can integrate out the fields  $\eta$ and $\tilde{\eta}$. Then the resulting action is
\begin{equation}\label{gfmulti}
S_b = \frac{1}{2} \int d^{q+1} \sigma\ \bigg( \partial_0 X^T g \partial_0 X  - \widetilde{\partial \! X}\tilde g  \widetilde{\partial\! X}\bigg) 
 - \int d^{q+1}\sigma\   \partial_0 X^T C \widetilde{\partial\! X} ,
\end{equation}
where
\bea
g\equiv g_{\mu\nu}, \qquad \tilde{g}\equiv \tilde{g}_{ij}, \qquad C\equiv C_{\mu i}.
\eea
We identify $g$, $\tilde{g}$ and $C$ as
\bea
g = \bigg(G^{-1} + \Pi \tilde G \Pi^T\bigg)^{-1}, \qquad \tilde g = \bigg(\tilde G^{-1} + \Pi^T  G \Pi\bigg)^{-1},\qquad
C =  - g \Pi \tilde G = - G \Pi \tilde g .
\label{membrane rel}
\eea
In the case of $q=1$, these relations are reduced to the closed-open string relations \eqref{close-open-string}. We rewrite the action after the Wick rotation ($\sigma^0\rightarrow -i\sigma^0$) in the compact matrix form
\begin{equation}\label{gfmulti}
S_{bE} = \frac{1}{2} \int d^{q+1}\sigma \ V^{\dagger} \begin{pmatrix} g & C \\ -C^T & \tilde{g} \end{pmatrix} V  ,
\end{equation}
where
\bea
\quad \exp(iS_b)=\exp(-S_{bE}), \qquad
V\equiv \left( \begin{array}{c} i\partial_0 X^{\mu} \\\widetilde{\partial\! X}^i \end{array} \right).
\eea
Let ${\cal G}$ denote the matrix
\bea
\cal G \equiv \begin{pmatrix} g & C \\ -C^T & \tilde{g} \end{pmatrix}.
\eea
The inverse matrix is given by
\bea
{\cal G}^{-1}=\begin{pmatrix} (g+C\tilde{g}^{-1}C^T)^{-1} & -(g+C\tilde{g}^{-1}C^T)^{-1}C\tilde{g}^{-1}
 \\ \tilde{g}^{-1}C^T(g+C\tilde{g}^{-1}C^T)^{-1} & (\tilde{g}+C^Tg^{-1}C)^{-1}  \end{pmatrix},
\eea
where we used the analytic inversion formula
\bea
\begin{pmatrix} a &b
 \\ c & d \end{pmatrix}^{-1}&=&\begin{pmatrix} a^{-1}+a^{-1}b(d-ca^{-1}b)^{-1}ca^{-1} &-a^{-1}b(d-ca^{-1}b)^{-1}
 \\ -(d-ca^{-1}b)^{-1}ca^{-1} & (d-ca^{-1}b)^{-1} \end{pmatrix}
 \nn\\
 &=&\begin{pmatrix} (a-bd^{-1}c)^{-1} &-(a-bd^{-1}c)^{-1}bd^{-1}
 \\ -d^{-1}c(a-bd^{-1}c)^{-1} & d^{-1}+d^{-1}c(a-bd^{-1}c)^{-1}bd^{-1} \end{pmatrix}.
\eea
We also have
\bea
{\cal H}&\equiv&  \begin{pmatrix} G & \Phi
 \\ -\Phi^T & \tilde{G} \end{pmatrix}^{-1}+ \begin{pmatrix} 0 & \Pi
 \\ -\Pi^T & 0 \end{pmatrix}
\nn\\ 
&=&\begin{pmatrix} (G+\Phi\tilde{G}^{-1}\Phi^T)^{-1} & -(G+\Phi\tilde{G}^{-1}\Phi^T)^{-1}\Phi\tilde{G}^{-1}+\Pi
 \\ \tilde{G}^{-1}\Phi^T(G+\Phi\tilde{G}^{-1}\Phi^T)^{-1}-\Pi^T & (\tilde{G}+\Phi^TG^{-1}\Phi)^{-1} \end{pmatrix}
 .
\eea
Interestingly, we can get the relations, which is similar to the closed-open string relations  by setting ${\cal G}^{-1}={\cal H}$. These relations are called generalized closed-open relations. These relations are
\bea
\label{gco1}
g+C\tilde{g}^{-1}C^T=G+\Phi\tilde{G}^{-1}\Phi^T,\qquad \tilde{g}+C^Tg^{-1}C=\tilde{G}+\Phi^T G^{-1}\Phi,
\eea
\bea
\label{gco2}
g^{-1}C=G^{-1}\Phi-\Pi\bigg(\tilde{G}+\Phi^TG^{-1}\Phi\bigg),\qquad \Phi\tilde{G}^{-1}=C\tilde{g}^{-1}+\bigg(g+C\tilde{g}^{-1}C^T\bigg)\Pi.
\eea
These relations imply that we can interchange
\bea
g\leftrightarrow G,\qquad \tilde{g}\leftrightarrow\tilde{G},\qquad C\leftrightarrow\Phi,\qquad \Pi\leftrightarrow -\Pi
\eea
to write the action in terms of $G$, $\Phi$ and $\Pi$. When $q=1$, we obtain
\bea
\frac{1}{g+B}=\frac{1}{G+\Phi}+\Pi.
\eea
We use ${\cal G}={\cal H}^{-1}$ to get another form of the generalized closed-open relations as well.
\bea
\begin{pmatrix} g & C
 \\ -C^T & \tilde{g} \end{pmatrix}={\cal H}^{-1}.
\eea
The results are
\bea
g^{-1}&=&\bigg(1-\Phi\Pi^T\bigg)^TG^{-1}\bigg(1-\Phi\Pi^T\bigg)+\Pi\tilde{G}\Pi^T,
\nn\\
\tilde{g}^{-1}&=&\bigg(1-\Phi^T\Pi\bigg)^T\tilde{G}^{-1}\bigg(1-\Phi^T\Pi\bigg)+\Pi^T G\Pi,
\nn\\
C&=&\bigg\lbrack\bigg(1-\Phi\Pi^T\bigg)^TG^{-1}\bigg(1-\Phi\Pi^T\bigg)+\Pi\tilde{G}\Pi^T\bigg\rbrack^{-1}\bigg\lbrack\bigg(1-\Phi\Pi^T\bigg)^TG^{-1}\Phi-\Pi\tilde{G}\bigg\rbrack.
\nn\\
\eea
Please refer to \ref{app1} for the detailed computations.

We can also use the generalized metric to derive the generalized closed-open relations. The generalized metric is exactly the matrix in the Hamiltonian. Starting from  
\begin{equation}\label{gfmulti}
S_{bE}  = \frac{1}{2} \int d^{q+1}\sigma\ V^{\dagger} \begin{pmatrix} g & C \\ -C^T & \tilde{g} \end{pmatrix} V  ,
\end{equation}
we get the Hamiltonian 
\bea
H[X, P]&=&\int d^q\sigma\ \bigg( \partial_0 X^T P-S_{bE}\bigg)
\nn\\
&=&\int d^q\sigma\ \bigg\lbrack \partial_0 X^T\bigg(g\partial_0X-iC\widetilde{\partial X}\bigg)-\frac{1}{2}\partial_0 X^T g\partial_0 X-\frac{1}{2}\widetilde{\partial X}^T\tilde{g}\widetilde{\partial X}+i\partial_0 X^TC\widetilde{\partial X}\bigg\rbrack
\nn\\
&=&\int d^q\sigma\ \bigg(\frac{1}{2}\partial_0X^T g\partial_0 X-\frac{1}{2}\widetilde{\partial X}^T\tilde{g}\widetilde{\partial X}\bigg)
\nn\\
&=&-\frac{1}{2}\int d^q\sigma\ \left( \begin{array}{c} iP \\\widetilde{\partial\! X}^i \end{array} \right)^T \begin{pmatrix} g^{-1} & -g^{-1}C \\ -C^Tg^{-1} & \tilde{g}+C^Tg^{-1}C \end{pmatrix}\left( \begin{array}{c} iP \\\widetilde{\partial\! X}^i \end{array} \right),
\eea
where
$P$ is the canonical momentum corresponding to the field $X$, i.e., $P=g\partial_0 X-iC\widetilde{\partial X}$. 
If we take $q=1$, the matrix in Hamiltonian is the usual generalized metric.

We can use another way to write the generalized metric
\bea
&&\begin{pmatrix} 1 & \Pi
 \\ 0 & 1 \end{pmatrix}\begin{pmatrix} 1 & 0
 \\ -\Phi^T & 1 \end{pmatrix}\begin{pmatrix} G^{-1} & 0
 \\ 0 & \tilde{G} \end{pmatrix}\begin{pmatrix} 1 & -\Phi
 \\ 0 & 1 \end{pmatrix}\begin{pmatrix} 1 & 0
 \\ \Pi^T & 1 \end{pmatrix}
 \nn\\
 &=&\begin{pmatrix} (1-\Pi\Phi^T)G^{-1}(1-\Phi\Pi^T)+\Pi\tilde{G}\Pi^T & -(1-\Pi\Phi^T)G^{-1}\Phi+\Pi\tilde{G}
 \\ -\Phi^TG^{-1}(1-\Phi\Pi^{T})+\tilde{G}\Pi^T & \Phi^TG^{-1}\Phi+\tilde{G} \end{pmatrix}
 \nn\\
 &=&\begin{pmatrix} g^{-1} & -g^{-1}C
 \\ -C^Tg^{-1} & \tilde{g}+C^Tg^{-1}C \end{pmatrix}.
\eea
We used \eqref{gco1} and \eqref{gco2} to get the second equality. In other words, we can get the generalized closed-open relations from the generalized metric. 

\subsection{Membrane Action}
Starting from the action
\begin{equation}\label{NambuGoto}
    S_M= -\int  d^{q+1}\sigma\ \sqrt{-\det{} (g_{\mu\nu} \partial_A X^\mu \partial_B X^\nu)} ,
\end{equation}
we introduce an auxiliary field $h_{AB}$ and write a classically equivalent action
\begin{equation}\label{membranesigma}
    S_{Mc} = -\frac{1}{2}\int d^{q+1} \sigma\ \sqrt{-\det{} h}\ \bigg(g_{\mu\nu} h^{AB} \partial_A X^\mu  \partial_B X^\nu -(q-1)\bigg) .
\end{equation}
We used an equation of motion of $h^{AB}$
\bea
\frac{1}{2}h_{AB}\bigg(h^{CD}\partial_{C}X^{\mu}\partial_D X^{\nu}g_{\mu\nu}-(q-1)\bigg)
=\partial_A X^{\mu}\partial_B X^{\nu} g_{\mu\nu}
\eea
to derive the equivalence. For $q\neq 1$, we have
\bea
h^{AB}\partial_A X^{\mu}\partial_B X^{\nu}g_{\mu\nu}=q+1.
\eea
Therefore, we get $h_{AB}=\partial_A X^{\mu}\partial_B X^{\nu} g_{\mu\nu}$. After fixing (by reparametrization invariance) the components $h_{a0}, h_{0b}$ and $h_{00}$ by choosing $h_{a0}= h_{0b}=0$ and $h_{00}= -\det (h_{a b})$, and using the equation of motion of $h^{ab}$
 \bea
 h_{ab}\bigg(h^{cd}\partial_c X^{\mu}\partial_d X^{\nu}g_{\mu\nu}-(q-1)\bigg)=\partial_a X^{\mu}\partial_b X^{\nu} g_{\mu\nu} ,
 \eea
 we get the classical equivalence with the gauge fixing
\bea
\label{gaugefixed}
    S_{\mbox{gf}}= \frac{1}{2} \int d^{q+1} \sigma\ \bigg\lbrack g_{\mu\nu} \partial_0 X^\mu \partial_0 X^\nu - \det{}\bigg(g_{\mu\nu} \partial_a X^\mu \partial_b X^\nu\bigg)\bigg\rbrack .
\eea
The action \eqref{gaugefixed} can be rewritten as
\begin{equation}\label{gfmulti}
S_{\mbox{gf}}= \frac{1}{2} \int d^{q+1}\sigma\ \bigg(  \partial_0 Xg \partial_0 X  -  \widetilde{\partial \! X} \tilde g\widetilde{\partial\! X}\bigg) .
\end{equation}
We can add a $(q+1)$-form background term, 
$\frac{1}{(q+1)!}C_{i_1i_2\cdots i_{q+1}} dx^{i_1} dx^{i_2} \cdots dx^{i_{q+1}}$. The action is
\begin{equation}\label{Cmulti}
    S_C = - \int d^{q+1}\sigma\ \partial_0 X C \widetilde{\partial\! X} .
\end{equation}
We combine $S_{gf}$ with $S_C$ to get the same action as the Nambu-Sigma model.

\subsection{Generalized DBI}
Before we generalize the DBI action, we first review the well-known theory, DBI theory, which is an effective action for an open string ending on a D-brane. The action is 
\bea 
\label{BIaction}
-\frac{1}{g_s}\int d^{p+1} x \sqrt{-\det{}( g +B+F)}
= -\frac{1}{g_s}\int d^{p+1}x \bigg( -\det{} g\bigg)^{\frac{1}{4}} \bigg\lbrack- \det{} \bigg( g - (B+F)g^{-1}(B+F)\bigg)\bigg\rbrack^{\frac{1}{4}},
\nn
\eea
where $g_s$, $g$ and $B$ are closed string coupling constant, metric and two-form antisymmetric background. $F$ is the abelian field strength ($F\equiv dA$). Before showing the equivalence between the commutative and non-commutative descriptions, we shall discuss the relations between the closed and open string parameters. These are
\bea
G_s=g_s\bigg(\frac{\det{}(G+\Phi)}{\det{}(g+B)}\bigg)^{\frac{1}{2}},
\eea
\bea
g-Bg^{-1}B=G-\Phi G^{-1}\Phi,\qquad
Bg^{-1}=\Phi G^{-1}-\bigg(G-\Phi G^{-1}\Phi\bigg)\Pi.
\eea
These relations imply that we can determine the open string variables from closed string variables by choosing $\Pi$. We can further rewrite $G_s$ as
\bea
G_s=g_s\bigg(\frac{\det{}(G+\Phi)}{\det{}(g+B)}\bigg)^{\frac{1}{2}}=g_s\bigg(\frac{\det{} G} {\det{} g}\bigg)^{\frac{1}{4}}\bigg(\frac{\det{}(G-\Phi G^{-1}\Phi)}{\det{}(g-B g^{-1}B)}\bigg)^{\frac{1}{4}}=g_s\bigg(\frac{\det{} G} {\det{} g}\bigg)^{\frac{1}{4}}.
\nn\\
\eea
Then we include the gauge field in the generalized metric
\bea
&&\begin{pmatrix} 1 & F
 \\ 0 & 1 \end{pmatrix}\begin{pmatrix} 1 & B
 \\ 0 & 1 \end{pmatrix}\begin{pmatrix} g & 0
 \\ 0 & g^{-1} \end{pmatrix}\begin{pmatrix} 1 & 0
 \\ -B & 1 \end{pmatrix}\begin{pmatrix} 1 & 0
 \\ -F & 1 \end{pmatrix}
 \nn\\
 &=&\begin{pmatrix}
  g-(B+F)g^{-1}(B+F) & (B+F)g^{-1}
 \\ -g^{-1}(B+F) & g^{-1} \end{pmatrix}
 \nn\\
 &=&\begin{pmatrix} 1 & F
 \\ 0 & 1 \end{pmatrix}\begin{pmatrix} 1 & 0
 \\ \Pi & 1 \end{pmatrix}\begin{pmatrix} 1 & \Phi
 \\ 0 & 1 \end{pmatrix}\begin{pmatrix} G & 0
 \\ 0 & G^{-1} \end{pmatrix}\begin{pmatrix} 1 & 0
 \\ -\Phi & 1 \end{pmatrix}\begin{pmatrix} 1 & -\Pi
 \\ 0 & 1 \end{pmatrix}\begin{pmatrix} 1 & 0
 \\ -F & 1 \end{pmatrix}
 \nn\\
&=&\begin{pmatrix} 1 & F
 \\ 0 & 1 \end{pmatrix}\begin{pmatrix} 1 & 0
 \\ \Pi & 1 \end{pmatrix}\begin{pmatrix} 1 & -F^{\prime}
 \\ 0 & 1 \end{pmatrix}\begin{pmatrix} 1 & (\Phi+F^{\prime})
 \\ 0 & 1 \end{pmatrix}\begin{pmatrix} G & 0
 \\ 0 & G^{-1} \end{pmatrix}
 \nn\\
 &&\cdot\begin{pmatrix} 1 & 0
 \\ -(\Phi+F^{\prime}) & 1 \end{pmatrix}\begin{pmatrix} 1 & 0
 \\ F^{\prime} & 1 \end{pmatrix}\begin{pmatrix} 1 & -\Pi
 \\ 0 & 1 \end{pmatrix}\begin{pmatrix} 1 & 0
 \\ -F & 1 \end{pmatrix}. \label{factorization}
\eea
We add one new block matrix $N$ to factorize the generalized metric. Later we will combine them to get the equivalence between the non-commutative and commutative descriptions.
\bea
\begin{pmatrix} 1 & 0
 \\ \Pi^{\prime} & 1 \end{pmatrix}\begin{pmatrix} N^T & 0
 \\ 0 & N^{-1} \end{pmatrix}\begin{pmatrix} 1 & \Phi^{\prime}
 \\ 0 & 1 \end{pmatrix}\begin{pmatrix} G & 0
 \\ 0 & G^{-1} \end{pmatrix}\begin{pmatrix} 1 & 0
 \\ -\Phi^{\prime} & 1 \end{pmatrix}\begin{pmatrix} N & 0
 \\ 0 & (N^{-1})^T \end{pmatrix}\begin{pmatrix} 1 & -\Pi^{\prime}
 \\ 0 & 1 \end{pmatrix},
\nn\\
\eea
where $\Phi^{\prime}$=$\Phi+F^{\prime}$.
From
\bea
&&\begin{pmatrix} 1 & 0
 \\ \Pi^{\prime} & 1 \end{pmatrix}\begin{pmatrix} N^T & 0
 \\ 0 & N^{-1} \end{pmatrix}=\begin{pmatrix} N^T & 0
 \\ \Pi^{\prime}N^T & N^{-1} \end{pmatrix}
 \nn\\
 &=&\begin{pmatrix} 1 & F
 \\ 0 & 1 \end{pmatrix}\begin{pmatrix} 1 & 0
 \\ \Pi & 1 \end{pmatrix}\begin{pmatrix} 1 & -F^{\prime}
 \\ 0 & 1 \end{pmatrix}
 =\begin{pmatrix} 1+F\Pi & -(1+F\Pi)F^{\prime}+F
 \\ \Pi & -\Pi F^{\prime}+1 \end{pmatrix},
\eea
we can obtain
\bea
\Pi^{\prime}&=&(1+\Pi F)^{-1}\Pi=\Pi(1+F\Pi)^{-1},
\nn\\
F^{\prime}&=&F(1+\Pi F)^{-1}=(1+F\Pi)^{-1}F,
\nn\\
N&=&1+\Pi F.
\eea
We find useful formulae from
\bea
&&\begin{pmatrix} 1 & 0
 \\ \Pi^{\prime} & 1 \end{pmatrix}\begin{pmatrix} N^T & 0
 \\ 0 & N^{-1} \end{pmatrix}\begin{pmatrix} 1 & \Phi^{\prime}
 \\ 0 & 1 \end{pmatrix}\begin{pmatrix} G & 0
 \\ 0 & G^{-1} \end{pmatrix}\begin{pmatrix} 1 & 0
 \\ -\Phi^{\prime} & 1 \end{pmatrix}\begin{pmatrix} N & 0
 \\ 0 & (N^{-1})^{T} \end{pmatrix}\begin{pmatrix} 1 & -\Pi^{\prime}
 \\ 0 & 1 \end{pmatrix}
 \nn\\
 &=&\begin{pmatrix}
  g-(B+F)g^{-1}(B+F) & (B+F)g^{-1}
 \\ -g^{-1}(B+F) & g^{-1} \end{pmatrix}.
\eea
Hence, we obtain
\bea
g-(B+F)g^{-1}(B+F)&=&N^T\bigg(G-\Phi^{\prime}G^{-1}\Phi^{\prime}\bigg)N,
\nn\\
(B+F)g^{-1}&=&-N^T\bigg(G-\Phi^{\prime}G^{-1}\Phi^{\prime}\bigg)N\Pi^{\prime}+N^T\Phi^{\prime}G^{-1}(N^T)^{-1},
\nn\\
g^{-1}&=&-\Pi^{\prime}N^T GN\Pi^{\prime}+\bigg(\Pi^{\prime}N^T\Phi^{\prime}+N^{-1}\bigg)G^{-1}
 \bigg(\Phi^{\prime}N\Pi^{\prime}+(N^T)^{-1}\bigg).
\nn
\eea
Thus, we have
\bea
\det{}\bigg(g-(B+F)g^{-1}(B+F)\bigg)&=&\det{}^2 {}(N)\det{}\bigg(G-\Phi^{\prime}G^{-1}\Phi^{\prime}\bigg),
\nn\\
(g+B+F)^{-1}&=&\Pi^{\prime}+\bigg(N^T(G+\Phi^{\prime})N\bigg)^{-1}.
\eea
The DBI action can be rewritten in terms of the closed string parameters instead of the open string parameters by using the above relations:
\bea
&&-\frac{1}{g_s}\bigg(-\det{} (g+B+F) \bigg)^{\frac{1}{2}}
\nn\\
&=&-\frac{1}{g_s}\bigg( -\det{} g\bigg)^{\frac{1}{4}} \bigg\lbrack-\det{} \bigg( g-(B+F)g^{-1}(B+F) \bigg) \bigg\rbrack^{\frac{1}{4}}
\nn\\
&=&-\frac{1}{g_s}\bigg( -\det{} g\bigg)^{\frac{1}{4}}\det{}^{\frac{1}{2}}\bigg(1+\Pi F\bigg) \bigg\lbrack -\det{}\bigg(G-\Phi^{\prime}G^{-1}\Phi^{\prime} \bigg) \bigg\rbrack^{\frac{1}{4}}
\nn\\
&=&-\frac{1}{G_s}\det{}^{\frac{1}{2}}\bigg(1+\Pi F\bigg)\bigg( -\det{}G\bigg)^{\frac{1}{4}}\bigg\lbrack -\det{} \bigg(G-\Phi^{\prime}G^{-1}\Phi^{\prime}\bigg) \bigg\rbrack^{\frac{1}{4}}
\nn\\
&=&-\frac{1}{G_s}\det{}^{\frac{1}{2}}\bigg(1+\Pi F\bigg) \bigg\lbrack-\det{}\bigg(G+\Phi^{\prime}\bigg)\bigg\rbrack^{\frac{1}{2}}.
\eea
Then we perform the Seiberg-Witten map to get
\bea
-\int d^{p+1}x\ \frac{1}{g_s}\bigg(-\det{}(g+B+F) \bigg)^{\frac{1}{2}}=-\int d^{p+1}\hat{x}\frac{1}{\hat{G}_s}\det{}^{\frac{1}{2}}\bigg(\frac{\hat{\Pi}}{\Pi}\bigg)\bigg\lbrack-\det{}\bigg(\hat{G}+\hat{\Phi}^{\prime}\bigg)\bigg\rbrack^{\frac{1}{2}},
\nn\\
\eea
where the superscript $\ \hat{}\ $ means that the fields are evaluated at the {\it covariant coordinates}, which comes from the change of coordinates, $x \mapsto \rho_A^*(x)=\hat x =  x + \Pi \hat A\ $ induced by a map $\Pi \mapsto \Pi' = (1+\Pi F)^{-1}\Pi$. The coordinate $\hat x^\mu$ is called {\it covariant coordinate}. We used 
\bea
\det{}(1+\Pi F)=\det{}\bigg(\frac{\hat{\Pi}}{\Pi}\bigg)\det{}^2\bigg(\frac{\partial x}{\partial\hat{x}}\bigg)
\eea
to establish the equivalence between the non-commutative and commutative descriptions in the DBI theory.

We expect the generalization of the DBI theory can be done in the similar construction. The generalized DBI action is first proposed in \cite{Jurco:2012yv}. They use the equivalence between the non-commutative and commutative descriptions to construct the generalized DBI theory. In the generalized DBI theory, the antisymmetric background field is $(q+1)$-form rather than the 2-form antisymmetric background field in the DBI theory. When $q$=1, the generalized DBI action is naturally reduced to the DBI theory. 

We include the details of the calculations in \ref{app2}. Among these, the two following formulae are crucial to determine the generalized DBI,
\begin{align}
& \det{}\bigg(g+(C+H)\tilde{g}^{-1}(C+H)^T\bigg)=\det{}^2\bigg(1-H\Pi^T\bigg)\det\bigg(G+\Phi^{\prime}\tilde{G}^{-1}\Phi^{\prime T}\bigg). \label{term1} \\ 
&\det{}\bigg(1-\Pi^T H\bigg)=\det{}\bigg(1-H\Pi^T\bigg)=\det{}\bigg(\frac{\hat{\Pi}}{\Pi}\bigg)\det{}^{q+1}\bigg(\frac{\partial x}{\partial\hat{x}}\bigg).
\end{align}

It is obvious that if there is a term like \eqref{term1}, we would get a term $\bigg(\det{}\frac{\partial x}{\partial \hat{x}}\bigg)^{2(q+1)}$ in the action. From this term, it can be postulated that the action can be
\bea
\label{pDBI}
S_{GDBI} =  -\int d^{p+1} x \ \frac{1}{g_b} \bigg(-\det {}g\bigg)^{\frac{q}{2(q+1)}}  \bigg\lbrack- \det{}\bigg(g +(C+H) \tilde g^{-1} (C+H)^T \bigg)\bigg\rbrack^{\frac{1}{2(q+1)}} 
\nn\\
\eea
because the term $\bigg(\det{}\frac{\partial x}{\partial \hat{x}}\bigg)^{2(q+1)}$ cancels with the Jacobian which arises from coordinate transformation, such that the Lagrangian is an integral density.
The coupling constant $g_b$ is called closed brane coupling constant. We can also rewrite the open brane coupling constant $G_b$ as
\bea
G_b=g_b\bigg(\frac{\det{} G} {\det{} g}\bigg)^{\frac{q}{2(q+1)}}\bigg(\frac{\det{}(G+\Phi \tilde{G}^{-1}\Phi^T)}{\det{}(g+C \tilde{g}^{-1}C^T)}\bigg)^{\frac{1}{2(q+1)}}=g_b\bigg(\frac{\det{} G} {\det{} g}\bigg)^{\frac{q}{2(q+1)}}.
\eea
We used
\bea
G+\Phi\tilde{G}^{-1}\Phi^T=g+C\tilde{g}^{-1}C^T
\eea
in the last equality.
The action of the generalized DBI can be rewritten by using the open brane parameters.
\bea
&&-\int d^{p+1} x \ \frac{1}{g_b} \bigg(-\det{}g\bigg)^{\frac{q}{2(q+1)}}  \bigg\lbrack- \det{}\bigg(g +(C+H) \tilde g^{-1} (C+H)^T \bigg)\bigg\rbrack^{\frac{1}{2(q+1)}}
\nn\\
&=&-\int d^{p+1} x \ \frac{1}{G_b} \bigg(-\det {}G\bigg)^{\frac{q}{2(q+1)}}  \bigg\lbrack- \det{}\bigg(g +(C+H) \tilde g^{-1} (C+H)^T \bigg)\bigg\rbrack^{\frac{1}{2(q+1)}}
\nn\\
&=&-\int d^{p+1}x\ \frac{1}{G_b} \bigg(-\det {}G\bigg)^{\frac{q}{2(q+1)}}\det{}^{\frac{1}{(q+1)}}\bigg(1-H\Pi^T \bigg)\bigg\lbrack-\det{}\bigg(G+\Phi^{\prime}\tilde{G}^{-1}\Phi^{\prime T}\bigg)\bigg\rbrack^{\frac{1}{2(q+1)}}
\nn\\
&=&-\int d^{p+1}\hat{x}\ \frac{1}{\hat{G}_b} \bigg(-\det{} \hat{G}\bigg)^{\frac{q}{2(q+1)}}\det{}^{\frac{1}{(q+1)}}\bigg(\frac{\hat{\Pi}}{\Pi}\bigg)\bigg\lbrack-\det{}\bigg(\hat{G}+\hat{\Phi}^{\prime}\hat{\tilde{G}}^{-1}\hat{\Phi}^{\prime T}\bigg)\bigg\rbrack^{\frac{1}{2(q+1)}}.
\nn\\
\eea
This action is based on the equivalence between the non-commutative and commutative gauge theories. The
closed-open relations can be generalized from the generalized metric. On the other hand, it can also be derived from the Nambu-Sigma model. If we consider a 2-form background, it goes back to the DBI theory. If we choose a 3-form background and set $p$=5,
the action is
\bea
S_\text{$5$-DBI} &=& - \int d^{6} x \ \frac{1}{g_b} \sqrt{-\det{}g}\  \det{}^{\frac{1}{6}}\bigg(1 +g^{-1}(C+H) \tilde g^{-1} (C+H)^T \bigg) 
\nn\\
&\approx&-\int d^{6} x \ \frac{1}{g_b}\sqrt{-\det{} g}\  \bigg(1 +\frac{1}{3}\mbox{Tr} k-\frac{1}{6}\mbox{Tr}k^2+\frac{1}{18}(\mbox{Tr }k)^2+\cdots \bigg)^{\frac{1}{2}} ,
\eea
where $k^{\mu}{}_{\nu}=\frac{1}{2}(H+C)^{\mu\rho\sigma}(H+C)_{\nu\rho\sigma}$. This action is consistent with \cite{Cederwall:1997gg} up to the second order. The supersymmetric extension and other formulations of the membrane theory are in \cite{Lee:2010ey, Park:2008qe}.

\section{Consistency of the Dimensional Reduction}
\label{3}
In this section, we discuss the dimensional reduction of the action \eqref{pDBI} without scalar fields. We first show the dimensional reduction from $(q+1)$-$(p+1)$ to $q$-$p$. We only consider flat spacetime, constant antisymmetric background, and $(q+1)$-form gauge field in $(q+1)$-dimensional worldvolume directions (without a time direction) in $q$-$p$ system. In other words, we have two types worldvolume directions. We use the non-dotted Greek letters to indicate the worldvolume directions without the antisymmetric background field and the dotted Greek letters to indicate the worldvolume directions with the antisymmetric  background field.
For a consistent notation, we define $(\dot1, \dot2, \cdots, \dot{q})$ $\equiv$ ($p-q$, $p-q+1$,$\cdots$, $p-1$). The generalized DBI theory \eqref{pDBI} gives
\begin{align}
S_{q+1,p+1} &=- \int d^{p+2}x\ \frac{1}{g_b} \det{}^{\frac{1}{2(q+2)}}\bigg(\delta_A{}^B+ H_{Ai} \tilde{g}^{ij} H_{Cj} g^{CB}\bigg) \nn
\\
&=- \int d^{p+2}x\ \frac{1}{g_b} \exp\bigg\lbrack\frac{1}{2(q+2)} \mathrm{Tr} \ln \bigg(\delta_A{}^B+ H_{Ai} \tilde{g}^{ij} H_{Cj} g^{CB}\bigg) \bigg\rbrack  \nn
\\
&=- \int d^{p+2}x\ \frac{1}{g_b} \exp\Bigg\lbrack\frac{1}{2(q+2)} \mathrm{Tr} \bigg(\sum_{n=1}^\infty \frac{(-1)^{n+1}}{n}\big(H_{Ai} \tilde{g}^{ij} H_{Cj} g^{CB}\big)^n \bigg) \Bigg\rbrack \label{p+1_p'+1}.
\end{align}
We used
\bea
\det{}^x(I+M)=\exp\bigg(x\mbox{Tr} \ln(I+M)\bigg), \qquad\ln(1-x)=-\sum_{n=1}^\infty \frac{x^n}{n}
\eea
to get the second and third equalities respectively. Then we calculate $ H_{Ai}\tilde{g}^{ij} H_{Cj} g^{CB}$
\begin{align*}
H_{Ai}\tilde{g}^{ij} H_{Cj} g^{CB} &= \frac{1}{\bigg((q+1)!\bigg)^2}H_{AC_1\cdots C_{q+1}}\sum_{\pi \in \sigma_{q+1}}\sgn(\pi)\bigg(g^{C_{\pi(1)}D_1}g^{C_{\pi(2)}D_2}\cdots g^{C_{\pi(q+1)}D_{q+1}}\bigg)H_{ED_1\cdots D_{q+1}}g^{EB}
\\
&=\frac{1}{\bigg((q+1)!\bigg)^2}\sum_{\pi \in \sigma_{q+1}}\sgn(\pi)H_A{}^{D_{\pi(1)}\cdots D_{\pi(q+1)}}H^B{}_{D_1...D_{q+1}}
\\
&=\frac{1}{(q+1)!}H_A{}^{D_1\cdots D_{q+1}}H^B{}_{D_1\cdots D_{q+1}}
\\
&\equiv \frac{1}{(q+1)!}(H^2)_A{}^B.
\end{align*}
The non-zero components in $(H^2)_A{}^B$ are
\begin{equation}
(H^2)_{p-q}{}^{p-q}=(H^2)_{p-q+1}{}^{p-q+1}=\cdots=(H^2)_{p+1}{}^{p+1}=(q+1)!(H^2_{p-q,p-q+1,\cdots, p+1}).
\end{equation}
Substituting the result and taking trace in the action \eqref{p+1_p'+1}, we get
\begin{align}
S_{q+1,p+1} &=- \int d^{p+2}x\ \frac{1}{g_m} \exp\bigg(\frac{1}{2(q+2)}  \sum_{n=1}^\infty \frac{(-1)^{n+1}}{n} (q+2)  (H_{p-q,p-q+1,\cdots, p+1})^{2n}   \bigg) \nonumber
\\
\label{p+1_p'+1_result}
&=-\int d^{p+2}x\ \frac{1}{g_m} \sum_{m=0}^\infty \frac{1}{m!} \bigg( \frac{1}{2} \sum_{n=1}^\infty \frac{(-1)^{n+1}}{n}  ( H_{p-q,p-q+1,\cdots, p+1})^{2n}   \bigg)^m.
\end{align}
Now we discuss the consistent truncation before we do the dimensional reduction.
The equation of motion of the generalized DBI theory at the leading order is
\bea
\partial_AH^{Ai}=0.
\eea 
We first fix the gauge such that a equation of motion of the generalized DBI theory becomes the wave equation at the leading order. When we compactify one direction, it becomes periodic.. The solution is proportional to $\exp({i\frac{n}{R}}y)$, where $R$ is radius of the compact torus, $n$ is the number of modes and y is the compacted coordinate. This periodic function gives a mass term in the equation of motion. When we shrink the radius to zero, the non-zero modes give the infinite mass and decouple from our theory consistently. 

If we compactify one direction and shrink the radius to zero, the ($q$+1)-form field strength becomes the $q$-form field strength.
The expression \eqref{p+1_p'+1_result} simply becomes
\begin{equation}
\label{pp'_comp}
S_{q,p} =-
\int d^{p+1}x\ \frac{1}{g_m} \sum_{m=0}^\infty \frac{1}{m!} \Bigg( \frac{1}{2} \sum_{n=1}^\infty \frac{(-1)^{n+1}}{n}  ( H_{p-q,p-q+1,\cdots, p})^{2n}   \Bigg)^m.
\end{equation}
In conclusion, we start from a system of $(q+1)$-$(p+1)$, we can get an effective action for $q$-$p$ system by the dimensional reduction. 

We want to emphasize that this is {\it not} a trivial check because the $2(q+1)$-th root in the action is so far predicted based on the equivalence between non-commutative and commutative gauge theories. This calculation in this simple example shows that the non-trivial power of this theory is also supported by the dimensional reduction.

\section{Comments on Pull-Back}
\label{4}
If we further require that the generalized DBI model can be reduced from $(q+1)$-$(p+1)$ to $q$-$p$ with scalar fields (by pull-back), the generalized DBI theory \eqref{pDBI} needs to include an one-form gauge potential for the $U(1)$ gauge symmetry. In the non-commutative gauge theory, we have these similar systems \cite{Ho:2013opa, Ho:2013paa, Ma:2012hc}. We discuss the possibility to extend the theory via the dimensional reduction. In this section, we show that inclusion of the one-form gauge field up to $H^2$ should be possible.

\subsection{Scalar Field and Gauge Potential}
When a worldvolume direction is compactified, the component of the compactified direction of a gauge potential $A^I$ gives a scalar field $X^I$,
\begin{align}
\label{compact}
A^I &\rightarrow X^I,  \qquad
F^{AI} \rightarrow  \partial^A X^I.
\end{align} 
The scalar field $X^I$ correspond to the positions of a brane in the transverse directions.

We introduce a scalar field from the pull-back. In the static gauge and the case of flat spacetime, we have
\begin{equation}
\label{metric}
g_{AB}=\eta_{AB}+\del_A X^I \del_B X^I.
\end{equation}
The inverse of this metric is 
\bea
\label{in_metric}
g^{AB}=\eta^{AB}+ \sum_{n=1}^\infty (-1)^{n} \bigg(\del^A X^{I_1}\bigg) \bigg(\del_{D_1} X^{I_1}\bigg) \bigg(\del^{D_1} X^{I_2}\bigg)\cdots\bigg(\del_{D_{n-1}} X^{I_{n-1}}\bigg)\bigg( \del^{D_{n-1}} X^{I_n}\bigg)\bigg( \del^B X^{I_n}\bigg),
\nn\\
\eea
which indeed satisfies a condition $g_{AB}g^{BC}=\delta_A{}^C$. 
We define
\bea
 \omega^{AB}\equiv\sum_{n=1}^\infty (-1)^{n} \bigg(\del^A X^{I_1}\bigg)\bigg(\del_{D_1} X^{I_1}\bigg)\bigg(\del^{D_1} X^{I_2}\bigg)\cdots\bigg(\del_{D_{n-1}} X^{I_{n-1}}\bigg)\bigg( \del^{D_{n-1}} X^{I_n}\bigg)\bigg( \del^B X^{I_n}\bigg)
\nn\\
 \eea
  for convenience and it is symmetric under interchanging the indices, i.e., $\omega^{AB}=\omega^{BA}$.

\subsection{$(q+1)$-$(p+1)$$\rightarrow$ $q$-$p$}
We show that the effective action of a $q$-$p$ brane system without the one-form gauge potential can be deduced from the $(q+1)$-$(p+1)$ system up to $H^2$ order. 

We also assume that only $\dot{\alpha}$ components of $H$ are turned on. The action is 
\begin{align}
S_{q+1,p+1} &= -\int d^{p+2}x\ \frac{1}{g_b}\sqrt{-\det{} g}~ \det{}^{\frac{1}{2(q+2)}}\bigg(\delta_A{}^B+ H_{Ai} \tilde{g}^{ij} H_{Cj} g^{CB}\bigg) \nonumber
\\
\label{pp'_Sca}
&=- \int d^{p+2}x\ \frac{1}{g_b}\sqrt{-\det{} g}~ \exp \bigg[\frac{1}{2(q+2)} \mbox{Tr} \bigg(\sum_{n=1}^\infty \frac{(-1)^{n+1}}{n}\big( H_{Ai} \tilde{g}^{ij} H_{Cj} g^{CB}\big)^n \bigg)\bigg] .
\end{align}
For $n=1$, we can obtain
\bea
\label{n=1_sca}
&&\mbox{Tr} \bigg(  H_{Ai} \tilde{g}^{ij} H_{Cj} g^{CB} \bigg)
\nn\\
&=& H_{p-q, p-q+1,\cdots, p+1}^2\sum_{k=0}^{q+1} \sum_{\substack{\da_k, \db_k, \dg_k=p-q}}^{p+1}\frac{q+2}{k!(q+2-k)!}
\nn\\
&& \times\eps_{\da_1\da_2\cdots \da_k \dg_{k+1}\dg_{k+2}\cdots \dg_{q+2}}\eps_{\db_1\db_2\cdots \db_k \dg_{k+1}\dg_{k+2}\cdots \dg_{q+2}} \om^{\da_1 \db_1}\om^{\da_2\db_2}\cdots \om^{\da_k \db_k},
\eea
where $\eps_{\da_1\da_2\cdots \da_k \dg_{k+1}\dg_{k+2}\cdots \dg_{q+2}}$ and $\eps_{\db_1\db_2\cdots \db_k \dg_{k+1}\dg_{k+2}\cdots \dg_{q+2}}$ are Levi-Civita symbols. The factorials $k!$ and $(q+2-k)!$ are used to cancel the factor of overcounting such that the coefficients of each term in the summation is simply unity. The expression \eqref{pp'_Sca} becomes
\bea
\label{pp'_final}
S_{q+1,p+1} 
&=&- \int d^{p+1}x\ \frac{1}{g_b}\sqrt{-\det{} g}
\nn\\
&&\exp \bigg( \frac{1}{2}H_{p-q, p-q+1,\cdots,p+1}^2  \sum_{k=0}^{q+1} \sum_{\substack{\da_k, \db_k, \dg_k=p-q}}^{p+1}\frac{1}{k!(q+2-k)!} 
\nn\\
&&\times\eps_{\da_1\da_2\cdots \da_k \dg_{k+1}\dg_{k+2}\cdots \dg_{q+2}}\eps_{\db_1\db_2\cdots \db_k \dg_{k+1}\dg_{k+2}\cdots \dg_{q+2}} \om^{\da_1 \db_1}\om^{\da_2\db_2}\cdots \om^{\da_k \db_k}+\cdots\bigg) \,.
\nn\\
\eea
The factors $(q+2)$ in \eqref{pp'_Sca} and \eqref{n=1_sca} cancel out each other.
If we compactify one worldvolume direction with background, say $p+1$, and shrink the radius to zero, then all $\om^{\dot{\a}({p+1})}\,$ vanish. This is equivalent to excluding $p+1$ in the summation, that is
\begin{align}
\sum_{\substack{\{\da_k, \db_k, \dg_k=p-q\}}}^{p+1} &\rightarrow \sum_{\substack{\{\da_k, \db_k,\dg_k=p-q\}}}^{p} \nn
\\
\sum_{k=0}^{q+1}  &\rightarrow  \sum_{k=0}^{q}
\end{align}
On the other hand, the Levi-Civita symbols should be modified to
\begin{equation}
\eps_{\da_1\da_2\cdots \da_k \dg_{k+1}\dg_{k+2}\cdots \dg_{q+2}}\eps_{\db_1\db_2\cdots \db_k \dg_{k+1}\dg_{k+2}\cdots \dg_{q+2}} \rightarrow (q+2-k)\eps_{\da_1\da_2\cdots \da_k \dg_{k+1}\dg_{k+2}\cdots \dg_{q+1}}\eps_{\db_1\db_2\cdots \db_k \dg_{k+1}\dg_{k+2}\cdots \dg_{q+1}} .
\end{equation}
As a result, the action \eqref{pp'_final} becomes
\bea
S_{q,p} 
&=& -\int d^{p+1}x\ \frac{1}{g_b}\sqrt{-\det{} g}
\nn \\
&&\times\exp \bigg( \frac{1}{2}H_{p-q, p-q+1,\cdots,p}^2  \sum_{k=0}^{q} \sum_{\substack{\da_k, \db_k, \dg_k=p-q}}^{p}\frac{1}{k!(q+1-k)!} 
\nn\\
&&\times\eps_{\da_1\da_2\cdots \da_k \dg_{k+1}\dg_{k+2}\cdots \dg_{q+1}}\eps_{\db_1\db_2\cdots \db_k \dg_{k+1}\dg_{k+2}\cdots \dg_{q+1}} \om^{\da_1 \db_1}\om^{\da_2\db_2}\cdots \om^{\da_k \db_k}+\cdots\bigg) ,
\nn\\
\eea
which is exactly the action \eqref{pp'_final} with $q+1$ and $p+1$ replaced by $q$ and $p$ respectively. This calculation shows the possibility to include the one-form gauge field in the theory up to $H^2$ order.

\section{Conclusion}
\label{5}
The generalized DBI is aimed for describing a $q$-brane ending on a $p$-brane. The most non-trivial feature of this action is the 2($q$+1)-th root, which is predicted by the existence of the equivalence between the commutative and non-commutative descriptions of the $q$-$p$ system. In this paper, we showed that the non-trivial power of the generalized DBI action can be consistent with the dimensional reduction. The calculation provides more evidences to the relation between 2-5 and M2-M5 system. In addition, we also find the possibility to extend the theory by adding the one-form gauge field in the presence of scalar fields from the dimensional reduction. We leave the full understanding of the dimensional reduction for ($q$+1)-($p$+1)$\rightarrow$ $q$-$p$ to the future. We can, of course, consider dimensional reduction along a direction orthogonal to the worldvolume directions. However, in our simple consideration, the background is not modified under this kind of the dimensional reduction. This should be trivial in this case. This direction should be a starting point to address the interesting issue of the duality structure of the higher-form fields. Although it is not a general study, we do not have many studies in the higher-form fields.

The supersymmetric extension of this theory should be able to give a direct link between the supergravity and generalized DBI theory. With the perturbative calculation up to the second order, we can obtain a similar form for the M5-brane \cite{Cederwall:1997gg}, which already has a supersymmetric extension. Although it is hard to find the supersymmetric theory for the DBI-form theory, we should be able to perform perturbative calculation order by order in principle.

So far, the relation between the generalized DBI theory and generalized geometry is unclear. However, the key point is that the generalized DBI theory is constructed by the equivalence between the commutative and non-commutative gauge theories. This equivalence is further governed by the generalized metric, which is always an important element in the generalized geometry. We expect that a generalized geometrical structure for $q>1$ can be found. 

The most interesting extension should be the T-duality rule between the background fields. Of course, we still have the familiar Buscher\rq{}s rule for $q$=1 with different values of $p$. Certainly, study of the T-duality rule of the generalized DBI theory is a challenging and interesting problem. The T-duality of the generalized DBI theory should be interesting on higher dimensional field theories (larger than eleven dimensions).

Finally, we remark on one related direction--double field theory of the DBI. By now, we do not get any insight to put the one-form gauge field in the double field theory. This is still an open problem. The starting point is to find the gauge transformation related to the Courant bracket. This should offer an unique structure to constrain the DBI theory in the double field theory. One more interesting prospect related to the open string of the double field theory is to understand the string sigma model with the manifest Buscher's rule. It is a well-known fact that the DBI model is equivalent to the calculation of the one-loop $\beta$ function of the string sigma model. If we can include the strong constraints in the double field theory of the string sigma model, the one-loop $\beta$ function would be an important consistent check. Moreover, one-loop $\beta$ function of the Nambu-Sigma model is also an important problem. So far, we only used the generalized metric and equivalence between the commutative and non-commutative descriptions to understand the generalized DBI model. We expect that the one-loop $\beta$ function of the Nambu-Sigma model should lead to the generalized DBI theory.

\appendix

\section{The Generalized Closed-Open Relations}
\label{app1}
We determine $g$ explicitly by 
\bea
g^{-1}&=&\bigg(G+\Phi\tilde{G}^{-1}\Phi^T\bigg)^{-1}
\nn\\
&&-\bigg\lbrack-\bigg(G+\Phi\tilde{G}^{-1}\Phi^T\bigg)^{-1}\Phi\tilde{G}^{-1}\bigg(\tilde{G}+\Phi^TG^{-1}\Phi\bigg)+\Pi\bigg(\tilde{G}+\Phi^TG^{-1}\Phi\bigg)\bigg\rbrack
\nn\\
&&\times\bigg\lbrack\tilde{G}^{-1}\Phi^T
\bigg(G+\Phi\tilde{G}^{-1}\Phi^T\bigg)^{-1}-\Pi^T\bigg\rbrack
\nn\\
&=&\bigg(G+\Phi\tilde{G}^{-1}\Phi^T\bigg)^{-1}+\bigg(G+\Phi\tilde{G}^{-1}\Phi^T\bigg)^{-1}\Phi\tilde{G}^{-1}\bigg(\tilde{G}+\Phi^TG^{-1}\Phi\bigg)\tilde{G}^{-1}\Phi^T
\bigg(G+\Phi\tilde{G}^{-1}\Phi^T\bigg)^{-1}
\nn\\
&&-\bigg(G+\Phi\tilde{G}^{-1}\Phi^T\bigg)^{-1}\Phi\tilde{G}^{-1}\bigg(\tilde{G}+\Phi^TG^{-1}\Phi\bigg)\Pi^T
-\Pi\bigg(\tilde{G}+\Phi^TG^{-1}\Phi\bigg)\tilde{G}^{-1}\Phi^T\bigg(G+\Phi\tilde{G}^{-1}\Phi^T\bigg)^{-1}
\nn\\
&&+\Pi\bigg(\tilde{G}+\Phi^TG^{-1}\Phi\bigg)\Pi^T.
\eea
Before showing the explicit answer, we show the trick for simplifying the third and fourth terms. 
The third term is
\bea
&&-\bigg(G+\Phi\tilde{G}^{-1}\Phi^T\bigg)^{-1}\Phi\tilde{G}^{-1}\bigg(\tilde{G}+\Phi^TG^{-1}\Phi\bigg)\Pi^T
\nn\\
&=&-\bigg(G+\Phi\tilde{G}^{-1}\Phi^T\bigg)^{-1}\bigg(\Phi+\Phi\tilde{G}^{-1}\Phi^TG^{-1}\Phi\bigg)\Pi^T
\nn\\
&=&-\bigg\lbrack\bigg(G+\Phi\tilde{G}^{-1}\Phi^T\bigg)G^{-1}\Phi\bigg(G^{-1}\Phi\bigg)^{-1}\bigg\rbrack^{-1}
\bigg(\Phi+\Phi\tilde{G}^{-1}\Phi^TG^{-1}\Phi\bigg)\Pi^T
\nn\\
&=&-G^{-1}\Phi\Pi^T.
\eea
The fourth term is
\bea
&&-\Pi\bigg(\tilde{G}+\Phi^TG^{-1}\Phi\bigg)\tilde{G}^{-1}\Phi^T\bigg(G+\Phi\tilde{G}^{-1}\Phi^T\bigg)^{-1}
\nn\\
&=&-\Pi\bigg(\Phi^T+\Phi^TG^{-1}\Phi\tilde{G}^{-1}\Phi^T\bigg)\bigg(G+\Phi\tilde{G}^{-1}\Phi^T\bigg)^{-1}
\nn\\
&=&-\Pi\bigg(\Phi^T+\Phi^TG^{-1}\Phi\tilde{G}^{-1}\Phi^T\bigg)\bigg\lbrack\bigg(\Phi^TG^{-1})^{-1}\Phi^TG^{-1}\bigg(G+\Phi\tilde{G}^{-1}\Phi^T\bigg)\bigg\rbrack^{-1}
\nn\\
&=&-\Pi\Phi^TG^{-1}.
\eea
By using the same method, the first and second terms are
\bea
&&\bigg(G+\Phi\tilde{G}^{-1}\Phi^T\bigg)^{-1}+G^{-1}\Phi\tilde{G}^{-1}\Phi^T\bigg(G+\Phi\tilde{G}^{-1}
\Phi^T\bigg)^{-1}
\nn\\
&=&\bigg(1+G^{-1}\Phi\tilde{G}^{-1}\Phi^T\bigg)\bigg(G+\Phi\tilde{G}^{-1}\Phi^T\bigg)^{-1}
\nn\\
&=&G^{-1}.
\eea
We can see the explicit answer by combining all terms
\bea
g^{-1}=\bigg(1-\Phi\Pi^T\bigg)^TG^{-1}\bigg(1-\Phi\Pi^T\bigg)+\Pi\tilde{G}\Pi^T.
\eea
Then we can determine $C$
\bea
C&=&-\bigg\lbrack\bigg(1-\Phi\Pi^T\bigg)^TG^{-1}\bigg(1-\Phi\Pi^T\bigg)+\Pi\tilde{G}\Pi^T\bigg\rbrack^{-1}
\nn\\
&&\times\bigg\lbrack-\bigg(G+\Phi\tilde{G}^{-1}\Phi^T\bigg)^{-1}\Phi
\tilde{G}^{-1}+\Pi\bigg\rbrack\bigg(\tilde{G}+\Phi^TG^{-1}\Phi\bigg)
\nn\\
&=&\bigg\lbrack\bigg(1-\Phi\Pi^T\bigg)^TG^{-1}\bigg(1-\Phi\Pi^T\bigg)+\Pi\tilde{G}\Pi^T\bigg\rbrack^{-1}\bigg\lbrack\bigg(1-\Phi\Pi^T\bigg)^TG^{-1}\Phi-\Pi\tilde{G}\bigg\rbrack.
\nn\\
\eea
The expression for $\tilde{g}^{-1}$ can also be derived
\bea
\tilde{g}^{-1}&=&\bigg(\tilde{G}+\Phi^TG^{-1}\Phi\bigg)^{-1}
\nn\\
&&+\bigg\lbrack\tilde{G}^{-1}\Phi^T-\Pi^T
\bigg(G+\Phi\tilde{G}^{-1}\Phi^T\bigg)\bigg\rbrack\bigg\lbrack\bigg(G+\Phi\tilde{G}^{-1}\Phi^T\bigg)^{-1}\Phi\tilde{G}^{-1}-\Pi
\bigg\rbrack
\nn\\
&=&\bigg(\tilde{G}+\Phi^TG^{-1}\Phi\bigg)^{-1}+\tilde{G}^{-1}\Phi^T
\bigg(G+\Phi\tilde{G}^{-1}\Phi^T\bigg)^{-1}\Phi\tilde{G}^{-1}
\nn\\
&&-\tilde{G}^{-1}\Phi^T\Pi
-\Pi^T\Phi\tilde{G}^T+\Pi^T\bigg(G+\Phi\tilde{G}^{-1}\Phi^T\bigg)\Pi.
\eea
The first term can be rewritten by using the fact that
\bea
\bigg(a+b\bigg)^{-1}=a^{-1}-a^{-1}b\bigg(b+ba^{-1}b\bigg)^{-1}ba^{-1}.
\eea
The above formula can be derived from the Binomial Inverse Theorem.
The first term is
\bea
&&\tilde{G}^{-1}-\tilde{G}^{-1}\bigg(\Phi^TG^{-1}\Phi\bigg)\bigg(\Phi^TG^{-1}\Phi+\Phi^TG^{-1}\Phi\tilde{G}^{-1}\Phi^TG^{-1}\Phi\bigg)^{-1}\Phi^TG^{-1}
\Phi\tilde{G}^{-1}
\nn\\
&=&\tilde{G}^{-1}-\tilde{G}^{-1}\Phi^T\bigg(\Phi^T+\Phi^TG^{-1}\Phi\tilde{G}^{-1}\Phi^T\bigg)^{-1}
\Phi^TG^{-1}\Phi\tilde{G}^{-1}
\nn\\
&=&\tilde{G}^{-1}-\tilde{G}^{-1}\Phi^T\bigg(1+G^{-1}\Phi\tilde{G}^{-1}\Phi^T\bigg)^{-1}
G^{-1}\Phi\tilde{G}^{-1}
\nn\\
&=&\tilde{G}^{-1}-\tilde{G}^{-1}\Phi^T\bigg(G+\Phi\tilde{G}^{-1}\Phi^T\bigg)^{-1}\Phi
\tilde{G}^{-1}.
\eea
If we combine the first term and second terms, we obtain $\tilde{G}^{-1}$. We can combine all terms to derive
\bea
\tilde{g}^{-1}=(1-\Phi^T\Pi)^T\tilde{G}^{-1}(1-\Phi^T\Pi)+\Pi^T G\Pi.
\eea

\section{Calculations for the Construction of the Generalized DBI}
\label{app2}
The generalization of DBI can also be done by a similar decomposition of matrix as \eqref{factorization}
\bea
&&\begin{pmatrix} 1 & -H^T
 \\ 0 & 1 \end{pmatrix}\begin{pmatrix} 1 & -C^T
 \\ 0 & 1 \end{pmatrix}\begin{pmatrix} \tilde{g} & 0
 \\ 0 & g^{-1} \end{pmatrix}\begin{pmatrix} 1 & 0
 \\ -C & 1 \end{pmatrix}\begin{pmatrix} 1 & 0
 \\ -H & 1 \end{pmatrix}
 \nn\\
 &=&\begin{pmatrix}
  \tilde{g}+(C+H)^Tg^{-1}(C+H) & -(C+H)^Tg^{-1}
 \\ -g^{-1}(C+H) & g^{-1} \end{pmatrix}
 \nn\\
 &=&\begin{pmatrix} 1 & -H^T
 \\ 0 & 1 \end{pmatrix}\begin{pmatrix} 1 & 0
 \\ \Pi & 1 \end{pmatrix}\begin{pmatrix} 1 & -\Phi^T
 \\ 0 & 1 \end{pmatrix}\begin{pmatrix} \tilde{G} & 0
 \\ 0 & G^{-1} \end{pmatrix}\begin{pmatrix} 1 & 0
 \\ -\Phi & 1 \end{pmatrix}\begin{pmatrix} 1 & \Pi^T
 \\ 0 & 1 \end{pmatrix}\begin{pmatrix} 1 & 0
 \\ -H & 1 \end{pmatrix}
 \nn\\
&=&\begin{pmatrix} 1 & -H^T
 \\ 0 & 1 \end{pmatrix}\begin{pmatrix} 1 & 0
 \\ \Pi & 1 \end{pmatrix}\begin{pmatrix} 1 & H^{\prime T}
 \\ 0 & 1 \end{pmatrix}\begin{pmatrix} 1 & -(\Phi^T+H^{\prime T})
 \\ 0 & 1 \end{pmatrix}\begin{pmatrix} \tilde{G} & 0
 \\ 0 & G^{-1} \end{pmatrix}
 \nn\\
 &&\cdot\begin{pmatrix} 1 & 0
 \\ -(\Phi^T+H^{\prime T}) & 1 \end{pmatrix}\begin{pmatrix} 1 & 0
 \\ H^{\prime} & 1 \end{pmatrix}\begin{pmatrix} 1 & \Pi^T
 \\ 0 & 1 \end{pmatrix}\begin{pmatrix} 1 & 0
 \\ -H & 1 \end{pmatrix}.
\eea 
We can add square matrix $M$ and $N$ in a similar way
\bea
\begin{pmatrix} 1 & 0
 \\ \Pi^{\prime} & 1 \end{pmatrix}\begin{pmatrix} N^T & 0
 \\ 0 & M^T \end{pmatrix}\begin{pmatrix} 1 & -\Phi^{\prime T}
 \\ 0 & 1 \end{pmatrix}\begin{pmatrix} \tilde{G} & 0
 \\ 0 & G^{-1} \end{pmatrix}\begin{pmatrix} 1 & 0
 \\ -\Phi^{\prime} & 1 \end{pmatrix}\begin{pmatrix} N & 0
 \\ 0 & M \end{pmatrix}\begin{pmatrix} 1 & \Pi^{\prime T}
 \\ 0 & 1 \end{pmatrix},
\eea
where $\Phi^{\prime}$=$\Phi+H^{\prime}$.
From
\bea
&&\begin{pmatrix} 1 & 0
 \\ \Pi^{\prime} & 1 \end{pmatrix}\begin{pmatrix} N^T & 0
 \\ 0 & M^T \end{pmatrix}=\begin{pmatrix} N^T & 0
 \\ \Pi^{\prime}N^T & M^T \end{pmatrix}
 \nn\\
 &=&\begin{pmatrix} 1 & -H^T
 \\ 0 & 1 \end{pmatrix}\begin{pmatrix} 1 & 0
 \\ \Pi & 1 \end{pmatrix}\begin{pmatrix} 1 & H^{\prime T}
 \\ 0 & 1 \end{pmatrix}
 =\begin{pmatrix} 1-H^T\Pi & (1-H^T\Pi)H^{\prime T}-H^T
 \\ \Pi & \Pi H^{\prime T}+1 \end{pmatrix},
\nn\\
\eea
we can obtain
\bea
\Pi^{\prime}&=&\Pi\bigg(1-H^T\Pi\bigg)^{-1},
\nn\\
H^{\prime}&=&H\bigg(1-\Pi^T H\bigg)^{-1},
\nn\\
N&=&1-\Pi^T H=\bigg(1+\Pi^{\prime T}H\bigg)^{-1},
\nn\\
M&=&1+H^{\prime}\Pi^T=\bigg(1-H\Pi^T\bigg)^{-1}.
\eea
We find the useful formulae from
\bea
&&\begin{pmatrix} 1 & 0
 \\ \Pi^{\prime} & 1 \end{pmatrix}\begin{pmatrix} N^T & 0
 \\ 0 & M^T \end{pmatrix}\begin{pmatrix} 1 & -\Phi^{\prime T}
 \\ 0 & 1 \end{pmatrix}\begin{pmatrix} \tilde{G} & 0
 \\ 0 & G^{-1} \end{pmatrix}\begin{pmatrix} 1 & 0
 \\ -\Phi^{\prime} & 1 \end{pmatrix}\begin{pmatrix} N & 0
 \\ 0 & M \end{pmatrix}\begin{pmatrix} 1 & \Pi^{\prime T}
 \\ 0 & 1 \end{pmatrix}
 \nn\\
 &=&\begin{pmatrix}
  \tilde{g}+(C+H)^T g^{-1}(C+H) & -(C+H)^Tg^{-1}
 \\ -g^{-1}(C+H) & g^{-1} \end{pmatrix}.
\eea
Therefore, we get
\bea
\tilde{g}+(C+H)^Tg^{-1}(C+H)&=&N^T\bigg(\tilde{G}+\Phi^{\prime T}G^{-1}\Phi^{\prime}\bigg)N,
\nn\\
-(C+H)^Tg^{-1}&=&N^T\bigg(\tilde{G}+\Phi^{\prime}G^{-1}\Phi^{\prime}\bigg)N
\Pi^{\prime T}-N^T\Phi^{\prime T}G^{-1}M,
\nn\\
g^{-1}&=&\Pi^{\prime}N^T \tilde{G}N\Pi^{\prime T}+\bigg(-\Pi^{\prime}N^T\Phi^{\prime T}+M^T\bigg)G^{-1}
 \bigg(-\Phi^{\prime}N\Pi^{\prime T}+M\bigg).
\nn
\eea
Thus, we have
\bea
\det{}\bigg(\tilde{g}+(C+H)^Tg^{-1}(C+H)\bigg)&=&\det{}^2 (N)\det{}\bigg(\tilde{G}+\Phi^{\prime T}G^{-1}\Phi^{\prime}\bigg)
\nn\\
&=&\det{}^2(1-\Pi^T H)\bigg(\tilde{G}+\Phi^{\prime T}G^{-1}\Phi^{\prime}\bigg).
\eea
From the bottom right block of
\bea
&&\begin{pmatrix} 1 & -\Pi^{\prime T}
 \\ 0 & 1 \end{pmatrix}\begin{pmatrix} N^{-1} & 0
 \\ 0 & M^{-1} \end{pmatrix}\begin{pmatrix} 1 & 0
 \\ \Phi^{\prime} & 1 \end{pmatrix}\begin{pmatrix} \tilde{G}^{-1} & 0
 \\ 0 & G \end{pmatrix}\begin{pmatrix} 1 & \Phi^{\prime T}
 \\ 0 & 1 \end{pmatrix}\begin{pmatrix} (N^T)^{-1} & 0
 \\ 0 & (M^T)^{-1} \end{pmatrix}\begin{pmatrix} 1 & 0
 \\ -\Pi^{\prime} & 1 \end{pmatrix}
 \nn\\
 &=&\begin{pmatrix}
  \tilde{g}+(C+H)^T g^{-1}(C+H) & -(C+H)^Tg^{-1}
 \\ -g^{-1}(C+H) & g^{-1} \end{pmatrix}^{-1},
\eea
we obtain
\bea
\det{}\bigg(g+(C+H)\tilde{g}^{-1}(C+H)^T\bigg)&=&\det{}\bigg\lbrack M^{-1}\bigg(G+\Phi^{\prime}\tilde{G}^{-1}\Phi^{\prime T}\bigg)\bigg(M^{-1}\bigg)^T\bigg\rbrack
\nn\\
&=&
\det{}^2\bigg(1-H\Pi^T\bigg)\det{}\bigg(G+\Phi^{\prime}\tilde{G}^{-1}\Phi^{\prime T}\bigg). 
\nn\\
\eea
In addition, we also have
\bea
\det{}\bigg(1-\Pi^T H\bigg)=\det{}\bigg(1-H\Pi^T\bigg)=\det{}\bigg(\frac{\hat{\Pi}}{\Pi}\bigg)\det{}^{q+1}\bigg(\frac{\partial x}{\partial\hat{x}}\bigg).
\eea

\section*{Acknowledgement}

We would like to thank Chi-Ming Chang, Song He, Pei-Ming Ho, Branislav Jurco, Pei-Wen Peggy Kao, Feng-Li Lin, Hisayoshi Muraki, Yen Chin Ong, Yiwen Pan, Peter Schupp, Yang Sun, Satoshi Watamura and You Wu for useful discussions.  This work is supported in part by NTU (grant \#NTU-CDP-
102R7708), National Science Council (grant \#101-2112-M-002-027-MY3), CASTS (grant \#103R891003) Taiwan, R.O.C..


\vskip .8cm
\baselineskip 22pt
\begin{thebibliography}{99}
\itemsep 0pt
\bibitem{Zwiebach:1992ie} 
  B.~Zwiebach,
  ``Closed string field theory: Quantum action and the B-V master equation,''
  Nucl.\ Phys.\ B {\bf 390}, 33 (1993)
  [hep-th/9206084].

\bibitem{Saadi:1989tb} 
  M.~Saadi and B.~Zwiebach,
  ``Closed String Field Theory from Polyhedra,''
  Annals Phys.\  {\bf 192}, 213 (1989).

\bibitem{Zwiebach:1985uq} 
  B.~Zwiebach,
  ``Curvature Squared Terms and String Theories,''
  Phys.\ Lett.\ B {\bf 156}, 315 (1985).

\bibitem{Hull:2009mi} 
  C.~Hull and B.~Zwiebach,
  ``Double Field Theory,''
  JHEP {\bf 0909}, 099 (2009)
  [arXiv:0904.4664 [hep-th]].

\bibitem{Hull:2009zb} 
  C.~Hull and B.~Zwiebach,
  ``The Gauge algebra of double field theory and Courant brackets,''
  JHEP {\bf 0909}, 090 (2009)
  [arXiv:0908.1792 [hep-th]].

\bibitem{Hohm:2010jy} 
  O.~Hohm, C.~Hull and B.~Zwiebach,
  ``Background independent action for double field theory,''
  JHEP {\bf 1007}, 016 (2010)
  [arXiv:1003.5027 [hep-th]].

\bibitem{Hohm:2010pp} 
  O.~Hohm, C.~Hull and B.~Zwiebach,
  ``Generalized metric formulation of double field theory,''
  JHEP {\bf 1008}, 008 (2010)
  [arXiv:1006.4823 [hep-th]].

\bibitem{Lee:2013hma} 
  K.~Lee and J.~H.~Park,
  ``Covariant action for a string in "doubled yet gauged" spacetime,''
  Nucl.\ Phys.\ B {\bf 880}, 134 (2014)
  [arXiv:1307.8377 [hep-th]].

\bibitem{Tseytlin:1990va} 
  A.~A.~Tseytlin,
  ``Duality symmetric closed string theory and interacting chiral scalars,''
  Nucl.\ Phys.\ B {\bf 350}, 395 (1991).

\bibitem{Tseytlin:1990nb} 
  A.~A.~Tseytlin,
  ``Duality Symmetric Formulation of String World Sheet Dynamics,''
  Phys.\ Lett.\ B {\bf 242}, 163 (1990).

\bibitem{Copland:2011wx} 
  N.~B.~Copland,
  ``A Double Sigma Model for Double Field Theory,''
  JHEP {\bf 1204}, 044 (2012)
  [arXiv:1111.1828 [hep-th]].

\bibitem{Berman:2007yf} 
  D.~S.~Berman and D.~C.~Thompson,
  ``Duality Symmetric Strings, Dilatons and O(d,d) Effective Actions,''
  Phys.\ Lett.\ B {\bf 662}, 279 (2008)
  [arXiv:0712.1121 [hep-th]].

\bibitem{Berman:2007xn} 
  D.~S.~Berman, N.~B.~Copland and D.~C.~Thompson,
  ``Background Field Equations for the Duality Symmetric String,''
  Nucl.\ Phys.\ B {\bf 791}, 175 (2008)
  [arXiv:0708.2267 [hep-th]].

\bibitem{Avramis:2009xi} 
  S.~D.~Avramis, J.~P.~Derendinger and N.~Prezas,
  ``Conformal chiral boson models on twisted doubled tori and non-geometric string vacua,''
  Nucl.\ Phys.\ B {\bf 827}, 281 (2010)
  [arXiv:0910.0431 [hep-th]].

\bibitem{Berkeley:2014nza} 
  J.~Berkeley, D.~S.~Berman and F.~J.~Rudolph,
  ``Strings and Branes are Waves,''
  JHEP {\bf 1406}, 006 (2014)
  [arXiv:1403.7198 [hep-th]].

\bibitem{Duff:1989tf} 
  M.~J.~Duff,
  ``Duality Rotations in String Theory,''
  Nucl.\ Phys.\ B {\bf 335}, 610 (1990).

\bibitem{Siegel:1993xq} 
  W.~Siegel,
  ``Two vierbein formalism for string inspired axionic gravity,''
  Phys.\ Rev.\ D {\bf 47}, 5453 (1993)
  [hep-th/9302036].

\bibitem{Siegel:1993th} 
  W.~Siegel,
  ``Superspace duality in low-energy superstrings,''
  Phys.\ Rev.\ D {\bf 48}, 2826 (1993)
  [hep-th/9305073].

\bibitem{Siegel:1993bj} 
  W.~Siegel,
  ``Manifest duality in low-energy superstrings,''
  In *Berkeley 1993, Proceedings, Strings '93* 353-363, and State U. New York Stony Brook - ITP-SB-93-050 (93,rec.Sep.) 11 p. (315661)
  [hep-th/9308133].

\bibitem{Gualtieri:2003dx} 
  M.~Gualtieri,
  ``Generalized complex geometry,''
  math/0401221 [math-dg].

\bibitem{Hitchin:2004ut} 
  N.~Hitchin,
  ``Generalized Calabi-Yau manifolds,''
  Quart.\ J.\ Math.\  {\bf 54}, 281 (2003)
  [math/0209099 [math-dg]].

\bibitem{Cavalcanti:2011wu} 
  G.~R.~Cavalcanti and M.~Gualtieri,
  ``Generalized complex geometry and T-duality,''
  A Celebration of the Mathematical Legacy of Raoul Bott (CRM Proceedings $\&$ Lecture Notes) American Mathematical Society (2010) 341-366. ISBN: 0821847775
  [arXiv:1106.1747 [math.DG]].

\bibitem{Hohm:2011dv} 
  O.~Hohm, S.~K.~Kwak and B.~Zwiebach,
  ``Double Field Theory of Type II Strings,''
  JHEP {\bf 1109}, 013 (2011)
  [arXiv:1107.0008 [hep-th]].

\bibitem{Hohm:2010xe} 
  O.~Hohm and S.~K.~Kwak,
  ``Frame-like Geometry of Double Field Theory,''
  J.\ Phys.\ A {\bf 44}, 085404 (2011)
  [arXiv:1011.4101 [hep-th]].

\bibitem{Aldazabal:2011nj} 
  G.~Aldazabal, W.~Baron, D.~Marques and C.~Nunez,
  ``The effective action of Double Field Theory,''
  JHEP {\bf 1111}, 052 (2011)
  [JHEP {\bf 1111}, 109 (2011)]
  [arXiv:1109.0290 [hep-th]].

\bibitem{Andriot:2012an} 
  D.~Andriot, O.~Hohm, M.~Larfors, D.~Lust and P.~Patalong,
  ``Non-Geometric Fluxes in Supergravity and Double Field Theory,''
  Fortsch.\ Phys.\  {\bf 60}, 1150 (2012)
  [arXiv:1204.1979 [hep-th]].

\bibitem{Geissbuhler:2013uka} 
  D.~Geissbuhler, D.~Marques, C.~Nunez and V.~Penas,
  ``Exploring Double Field Theory,''
  JHEP {\bf 1306}, 101 (2013)
  [arXiv:1304.1472 [hep-th]].

\bibitem{Grana:2012rr} 
  M.~Grana and D.~Marques,
  ``Gauged Double Field Theory,''
  JHEP {\bf 1204}, 020 (2012)
  [arXiv:1201.2924 [hep-th]].

\bibitem{Ma:2014ala} 
  C.~T.~Ma and C.~M.~Shen,
  ``Cosmological Implications from O(D,D),''
  Fortsch.\ Phys.\  {\bf 62}, 921 (2014)
  [arXiv:1405.4073 [hep-th]].

\bibitem{Hohm:2013jaa} 
  O.~Hohm, W.~Siegel and B.~Zwiebach,
  ``Doubled $\alpha'$-geometry,''
  JHEP {\bf 1402}, 065 (2014)
  [arXiv:1306.2970 [hep-th]].

\bibitem{Bedoya:2014pma} 
  O.~A.~Bedoya, D.~Marques and C.~Nunez,
  ``Heterotic $\alpha$'-corrections in Double Field Theory,''
  JHEP {\bf 1412}, 074 (2014)
  [arXiv:1407.0365 [hep-th]].

\bibitem{Berman:2010is} 
  D.~S.~Berman and M.~J.~Perry,
  ``Generalized Geometry and M theory,''
  JHEP {\bf 1106}, 074 (2011)
  [arXiv:1008.1763 [hep-th]].

\bibitem{Hohm:2013vpa} 
  O.~Hohm and H.~Samtleben,
  ``Exceptional Field Theory I: $E_{6(6)}$ covariant Form of M-Theory and Type IIB,''
  Phys.\ Rev.\ D {\bf 89}, no. 6, 066016 (2014)
  [arXiv:1312.0614 [hep-th]].

\bibitem{Hohm:2013uia} 
  O.~Hohm and H.~Samtleben,
  ``Exceptional field theory. II. E$_{7(7)}$,''
  Phys.\ Rev.\ D {\bf 89}, 066017 (2014)
  [arXiv:1312.4542 [hep-th]].

\bibitem{Hohm:2014fxa} 
  O.~Hohm and H.~Samtleben,
  ``Exceptional field theory. III. E$_{8(8)}$,''
  Phys.\ Rev.\ D {\bf 90}, 066002 (2014)
  [arXiv:1406.3348 [hep-th]].

\bibitem{Berman:2012vc} 
  D.~S.~Berman, M.~Cederwall, A.~Kleinschmidt and D.~C.~Thompson,
  ``The gauge structure of generalised diffeomorphisms,''
  JHEP {\bf 1301}, 064 (2013)
  [arXiv:1208.5884 [hep-th]].

\bibitem{Hohm:2013pua} 
  O.~Hohm and H.~Samtleben,
  ``Exceptional Form of D=11 Supergravity,''
  Phys.\ Rev.\ Lett.\  {\bf 111}, 231601 (2013)
  [arXiv:1308.1673 [hep-th]].

\bibitem{Hohm:2013bwa} 
  O.~Hohm, D.~Lüst and B.~Zwiebach,
  ``The Spacetime of Double Field Theory: Review, Remarks, and Outlook,''
  Fortsch.\ Phys.\  {\bf 61}, 926 (2013)
  [arXiv:1309.2977 [hep-th]].

\bibitem{Aldazabal:2013sca} 
  G.~Aldazabal, D.~Marques and C.~Nunez,
  ``Double Field Theory: A Pedagogical Review,''
  Class.\ Quant.\ Grav.\  {\bf 30}, 163001 (2013)
  [arXiv:1305.1907 [hep-th]].

\bibitem{Berman:2013eva} 
  D.~S.~Berman and D.~C.~Thompson,
  ``Duality Symmetric String and M-Theory,''
  Phys.\ Rept.\  {\bf 566}, 1 (2014)
  [arXiv:1306.2643 [hep-th]].

\bibitem{Ho:2008nn} 
  P.~M.~Ho and Y.~Matsuo,
  ``M5 from M2,''
  JHEP {\bf 0806}, 105 (2008)
  [arXiv:0804.3629 [hep-th]].

\bibitem{Ho:2012dn} 
  P.~M.~Ho, C.~T.~Ma and C.~H.~Yeh,
  ``BPS States on M5-brane in Large C-field Background,''
  JHEP {\bf 1208}, 076 (2012)
  [arXiv:1206.1467 [hep-th]].

\bibitem{Jurco:2012yv} 
  B.~Jurco and P.~Schupp,
  ``Nambu-Sigma model and effective membrane actions,''
  Phys.\ Lett.\ B {\bf 713}, 313 (2012)
  [arXiv:1203.2910 [hep-th]].

\bibitem{Schupp:2012nq} 
  P.~Schupp and B.~Jurco,
  ``Nambu Sigma Model and Branes,''
  PoS CORFU {\bf 2011}, 045 (2011)
  [arXiv:1205.2595 [hep-th]].

\bibitem{Jurco:2013upa} 
  B.~Jurco, P.~Schupp and J.~Vysoky,
  ``On the Generalized Geometry Origin of Noncommutative Gauge Theory,''
  JHEP {\bf 1307}, 126 (2013)
  [arXiv:1303.6096 [hep-th]].

\bibitem{Jurco:2014sda} 
  B.~Jurčo, P.~Schupp and J.~Vysoký,
  ``Extended generalized geometry and a DBI-type effective action for branes ending on branes,''
  JHEP {\bf 1408}, 170 (2014)
  [arXiv:1404.2795 [hep-th]].

\bibitem{Cederwall:1997gg} 
  M.~Cederwall, B.~E.~W.~Nilsson and P.~Sundell,
  ``An Action for the superfive-brane in D = 11 supergravity,''
  JHEP {\bf 9804}, 007 (1998)
  [hep-th/9712059].

\bibitem{Lee:2010ey} 
  K.~Lee and J.~H.~Park,
  ``Partonic description of a supersymmetric p-brane,''
  JHEP {\bf 1004}, 043 (2010)
  [arXiv:1001.4532 [hep-th]].

\bibitem{Park:2008qe} 
  J.~H.~Park and C.~Sochichiu,
  ``Taking off the square root of Nambu-Goto action and obtaining Filippov-Lie algebra gauge theory action,''
  Eur.\ Phys.\ J.\ C {\bf 64}, 161 (2009)
  [arXiv:0806.0335 [hep-th]].

\bibitem{Ho:2013opa} 
  P.~M.~Ho and C.~T.~Ma,
  ``S-Duality for D3-Brane in NS-NS and R-R Backgrounds,''
  JHEP {\bf 1411}, 142 (2014)
  [arXiv:1311.3393 [hep-th]].

\bibitem{Ho:2013paa} 
  P.~M.~Ho and C.~T.~Ma,
  ``Effective Action for Dp-Brane in Large RR (p-1)-Form Background,''
  JHEP {\bf 1305}, 056 (2013)
  [arXiv:1302.6919 [hep-th]].

\bibitem{Ma:2012hc} 
  C.~T.~Ma and C.~H.~Yeh,
  ``Supersymmetry and BPS States on D4-brane in Large C-field Background,''
  JHEP {\bf 1303}, 131 (2013)
  [arXiv:1210.4191 [hep-th]].

\end{thebibliography}
\end{document}